\documentclass[11pt,prd,aps,amssymb,amsmath,tightenlines,showpacs]{revtex4}
\usepackage{graphicx}
\usepackage{slashed}

\newcommand{\be}{\begin{equation}}
\newcommand{\ee}{\end{equation}}
\newcommand{\bea}{\begin{eqnarray}}
\newcommand{\eea}{\end{eqnarray}}
\newcommand{\nn}{\nonumber\\}

\newcommand{\half}{\textstyle{\frac{1}{2}}}
\newcommand{\cP}{{\cal P}}
\newcommand{\cT}{{\cal T}}
\newcommand{\cC}{{\cal C}}
\newcommand{\cH}{{\cal H}}
\newcommand{\cPT}{{\cal PT}}
\newcommand{\cCPT}{{\cal CPT}}

\begin{document}
 
\preprint[{\leftline{KCL-PH-TH/2015-{\bf 01}}}
\vspace{1.0cm}
\title{Foldy-Wouthuysen transformation for non-Hermitian Hamiltonians}

\author{Jean Alexandre$^1$}
\email{jean.alexandre@kcl.ac.uk}

\author{Carl M. Bender$^{2,1}$}
\email{cmb@wustl.edu}

\affiliation{\it $^1$Department of Physics, King's College London, WC2R 2LS, UK
\\ $^2$Department of Physics, Washington University, St. Louis, MO 63130, USA}

\date{\today}

\begin{abstract}
Two Non-Hermitian fermion models are proposed and analyzed by using
Foldy-Wouthuysen transformations. One model has Lorentz symmetry breaking and
the other has a non-Hermitian mass term. It is shown that each model has real
energies in a given region of parameter space, where they have a locally
conserved current.
\end{abstract} 

\pacs{11.30.Er, 03.65.-w, 03.70.+k}
\maketitle
 
\section{Introduction} 
\label{s1}
This work is concerned with fermion theories that are described by non-Hermitian
Hamiltonian densities. (For earlier studies, see Refs.~\cite{r1,r2,r3,r4}.) We
are particularly interested in fermion models that violate Lorentz invariance.

Our general procedure is to begin with a non-Hermitian Hamiltonian density of
the form $H=\psi^\dag\cH\psi$ for which the Hamiltonian operator $\cH$ has
eigenvalues $\omega$, which are assumed to be real in a region of parameter
space. The Schr\"odinger form of the equation of motion is
$$i\partial_0\psi=\cH\psi.$$
We then introduce a mapping $U$ to implement the Foldy-Wouthuysen (FW)
transformation \cite{r5} on the free fermion field $\psi$ and $\cH$, such that
$$\chi\equiv U\psi\quad\mbox{and}\quad U\cH U^{-1}\equiv\omega\gamma^0.$$
The equation of motion can then be rewritten as
$$i\partial_0\chi=\omega\gamma^0\chi,$$
where the FW Hamiltonian density is $H_{FW}=\chi^\dag\omega\gamma^0\chi=\omega
\bar\chi\chi$.

The motivation for this construction is to map the non-Hermitian Hamiltonian
density $H$ onto the FW Hamiltonian density $H_{FW}$, which is Hermitian (for
real $\omega$), with the same eigenvalues $\omega$ as the original Hamiltonian
operator $\cH$. Unlike the Dirac case, the operator $U$ in this paper is not
unitary, and thus $H_{FW}\neq H$. Indeed, 
\begin{equation}
H=\psi^\dag\cH\psi=\psi^\dag U^{-1}U\cH U^{-1}U\psi=\psi^\dag
U^{-1}\omega\gamma^0\chi\neq(U\psi)^\dag\omega\gamma^0\chi=H_{FW}.
\label{e1}
\end{equation}
In Ref.~\cite{r6} it is shown that the FW transformation for an extension of
the Dirac equation that contains all the Lorentz-symmetry and $\cCPT$-violating
terms is consistent with the Standard Model Extension (SME) \cite{r7}. However,
Ref.~\cite{r6} deals with Hermitian Lagrangians and does not overlap with the
present study. We also note that a non-Hermitian Lagrangian is studied in
Ref.~\cite{r8}, for which the corresponding Hamiltonian is Hermitian. This
feature is possible because of the presence of mixed derivatives $\partial_0
\partial_k$ in the Lagrangian.
 
\subsection{Derivation of the equation of motion}
Before proceeding, it is important to explain the derivation of the equation of
motion. We will show that the equation of motion may be obtained by the usual
(formal) procedure of performing a variational derivative of the action with
respect to $\bar\psi$ and not varying $\psi$. (Of course, this procedure is
questionable for a non-Hermitian Lagrangian because varying the action with
respect to $\psi$ does not reproduce the same equation of motion after taking
the Hermitian conjugate.) 
 
The four fermion components may be written as $\psi_a=\phi_a+i\chi_a$, where
$\phi_a$ and $\chi_a$ are real. The action then has the form
$$S=\int\bar\psi\left(i\slashed\partial-m-\Gamma\right)\psi=\int(\phi_b-i\chi_b)
\left(i\gamma^0\slashed\partial-m\gamma^0-\gamma^0\Gamma\right)_{bc}(\phi_c+
i\chi_c),$$
where $\Gamma$ satisfies $\Gamma^\dag=\Gamma$ and $\{\gamma^0,\Gamma\}=0$.

The equations of motion can then be correctly obtained by demanding that the
variational derivatives of $S$ with respect to $\phi_a$ and $\chi_a$
independently vanish. We obtain 
\bea
\frac{\delta S}{\delta\phi_a}
&=&\left(i\gamma^0\slashed\partial-m\gamma^0-\gamma^0\Gamma\right)_{ac}(\phi+
i\chi)_c-\left(-i\gamma^0\slashed\partial-m\gamma^0-\gamma^0\Gamma\right)_{
ba}(\phi-i\chi)_b\nonumber\\
&=&\left[i\gamma^0\slashed\partial+i\left(\gamma^0\slashed\partial\right)^T-m
\gamma^0+m\gamma^{0T}-\gamma^0\Gamma+(\gamma^0\Gamma)^T\right]_{ac}\phi_c\nn
&&+\left[-\gamma^0\slashed\partial+\left(\gamma^0\slashed\partial\right)^T-im
\gamma^0-im\gamma^{0T}-i\gamma^0\Gamma-i(\gamma^0\Gamma)^T\right]_{ac}\chi_c.
\nonumber
\eea
Then, from the properties of gamma matrices in the Dirac representation, we get
\bea
\frac{\delta S}{\delta\phi_a}
&=&2\left[i\gamma^0(\gamma^0\partial_0+\gamma^1\partial_1+\gamma^3\partial_3)-
\gamma^0\Gamma~\frac{1+\epsilon}{2}\right]_{ac}\phi_c\nonumber\\
&&+2\left[-\gamma^0\gamma^2\partial_2-im\gamma^0-i\gamma^0\Gamma~\frac{1-
\epsilon}{2}\right]_{ac}\chi_c,\nonumber
\eea
where $\epsilon=+1$ if $\Gamma$ is real and $\epsilon=-1$ if $\Gamma$ is
imaginary. Similar steps lead to
\bea
\frac{\delta S}{\delta\chi_a}&=&-2i\left[i\gamma^0\gamma^2\partial_2-m\gamma^0-
\gamma^0\Gamma~\frac{1-\epsilon}{2}\right]_{ac}\phi_c\nonumber\\
&&+2\left[i\gamma^0(\gamma^0\partial_0+\gamma^1\partial_1+\gamma^3\partial_3)-
\gamma^0\Gamma~\frac{1+\epsilon}{2}\right]_{ac}\chi_c.\nonumber
\eea
It is now easy to see that 
$$\frac{1}{2}\frac{\delta S}{\delta\phi_a}+\frac{i}{2}\frac{\delta S}{\delta
\chi_a}=\left(\gamma^0(i\slashed\partial-m-\Gamma)(\phi+i\chi)\right)_a,$$
which is independent of $\epsilon$. The latter equation is equivalent to
$\gamma^0\frac{\delta S}{\delta\psi^\star}=(i\slashed\partial-m-\Gamma)\psi$,
which agrees with the variation $\delta S/\delta\bar\psi$ performed with $\bar
\psi$ and $\psi$ considered as {\it independent} fields. Hence, we may derive
the equations of motion by following the formal conventional procedure.

\section{Model of Lorentz-symmetry violation} 
\label{s2} 
In this section we study a specific form of Lorentz-symmetry-violating
kinematics for $(3+1)$-dimensional fermions, which have real energies if the
parameters of the model satisfy certain inequalities. Using the metric 
$\eta_{\mu\nu}={\rm diag}(1,-1,-1,-1)$, we begin with the Lagrangian
\be
{\cal L}=\bar\psi\left(i\slashed\partial-i\slashed b-m\right)\psi,
\label{e2}
\ee
where $b_\mu$ represents the (real) vacuum expectation value of a vector and
this signals the breaking of Lorentz symmetry. The term $\bar\psi\slashed b\psi$
(without the factor of $i$) appears naturally in the SME. This term can be
obtained from the QED one-loop effective action in curved spacetime \cite{r9},
where $b_\mu\propto\frac{\alpha}{m^2}\partial_\mu R$, $\alpha$ is the
fine-structure constant, $m$ is the electron mass, and $R$ is the Ricci scalar
for the curved background.

The Hamiltonian density corresponding to the Lagrangian (\ref{e2}) is
$$H=\frac{{\cal L}\overleftarrow\partial}{\partial\dot\psi}\dot\psi-{\cal L}=
\bar\psi\left(i\vec\gamma\cdot\vec\nabla+i\slashed b+m\right)\psi,$$
where $\dot\psi=\partial_0\psi$. Note that $H$ is not Hermitian because it
contains the anti-Hermitian term $i\bar\psi\slashed b\psi=-\left(i\bar\psi
\slashed b\psi\right)^\dag$. The dispersion relation corresponding to the
Lagrangian (\ref{e2}) is 
$$(\omega-ib_0)^2=m^2+(\vec p-i\vec b)^2.$$
However, if we choose $b_0=0$ and restrict our attention to motion in the plane
perpendicular to $\vec b$, the energies $\omega$ satisfy
\be
\omega^2=m^2+p^2-b^2\quad{\rm for}\quad b_0=0=\vec p\cdot\vec b,
\label{e3}
\ee
and they are {\it real} for any momentum $\vec p$ as long as $b^2\le m^2$. The
existence of imaginary energies for low momentum (when $b^2>m^2$) implies the
possibility of runaway modes, which is a problem known in SME studies (see
Ref.~\cite{r10} for a recent discussion). (Runaway modes in $\cPT$-symmetric
theories are also considered in Ref.~\cite{r11}.)

\subsection{Current conservation}
To verify current conservation we multiply the equation of motion on the left by
$\bar\psi$ and then multiply the Hermitian conjugate of the equation of motion
on the right by $\gamma^0\psi$ gives
$$i\bar\psi(\slashed\partial-\vec b\cdot\vec\gamma)\psi=m\bar\psi\psi\quad{\rm
and}\quad-i\bar\psi(\overleftarrow{\slashed\partial}+\vec b\cdot\vec\gamma)\psi
=m\bar\psi\psi.$$
Next, we subtract these two equations and obtain
$$\partial_\mu(\bar\psi\gamma^\mu\psi)+2\bar\psi(\vec b\cdot\vec\gamma)\psi=0.$$

As usual, we define the probability density as $\rho\equiv\bar\psi\gamma^0\psi=
\psi^\dag\psi$ and the current density as $\vec j=\bar\psi\vec\gamma\psi$.
The previous equation now reads
\be
\dot\rho-\vec\nabla\cdot\vec j=-2\vec b\cdot\vec j.
\label{e4}
\ee
Because we restrict our attention to motion in the plane perpendicular to $\vec
b$, the current $\vec j$ is perpendicular to $\vec b$ and thus $\vec b\cdot\vec
j=0$. Therefore, the continuity equation (\ref{e4}) has the usual form
$\partial_\mu(\bar\psi\gamma^\mu\psi)=0$ and the current $\bar\psi\gamma^\mu
\psi$ is conserved.

\subsection{Charge-conjugation, parity, and time-reversal symmetries}
In the framework of the SME, the Lorentz-symmetry-violating term $\bar\psi~\vec
b\cdot\vec\gamma~\psi$ (without the factor of $i$), is $\cCPT$-odd \cite{R7}.
In our case, with $\vec b\to i\vec b$, we can check each symmetry independently:

$\bullet$ {\it Charge-conjugation} $\cC$:
$$C^{-1}(i\bar\psi~\vec b\cdot\vec\gamma~\psi)C=i\bar\psi({\cal C}^{-1}~\vec
b\cdot\vec\gamma^T~{\cal C})\psi=-i\bar\psi~\vec b\cdot\vec\gamma~\psi,$$
where $\cC$ satisfies $\cC^{-1}\gamma_\mu\cC=\cC\gamma_\mu\cC^{-1}=-
\gamma_\mu^T$, such that $\bar\psi~i\vec b\cdot\vec\gamma~\psi$ is $C$-odd.

$\bullet$ {\it Parity} $\cP$:
$$\cP^{-1}(i\bar\psi~\vec b\cdot\vec\gamma~\psi)\cP=i\bar\psi\gamma^0~\vec b
\cdot\vec\gamma~\gamma^0\psi=-i\bar\psi~\vec b\cdot\vec\gamma~\psi,$$
such that $\bar\psi~i\vec b\cdot\vec\gamma~\psi$ is $\cP$-odd.

$\bullet$ {\it Time-reversal} $\cT$:
$$\cT^{-1}(i\bar\psi~\vec b\cdot\vec\gamma~\psi)\cT=-i\bar\psi\gamma^5\cC^{-1}~
\vec b\cdot\vec\gamma^\star~\cC\gamma^5\psi=i\bar\psi\gamma^5~\vec b\cdot\vec
\gamma^\dag~\gamma^5\psi=i\bar\psi~\vec b\cdot\vec\gamma~\psi,$$
such that $\bar\psi~i\vec b\cdot\vec\gamma~\psi$ is $\cT$-even.

The non-Hermitian term $i\bar\psi~\vec b\cdot\vec\gamma~\psi$ is therefore even
under $\cCPT$ but odd under $\cPT$.

\subsection{Foldy-Wouthuysen transformation}
The Schr\"odinger form of the equation of motion obtained from the Lagrangian
(\ref{e2}) is
$$i\partial_0\psi=\cH\psi,$$
where the Hamiltonian operator is $\cH=\gamma^0\left[(\vec p-i\vec b)\cdot\vec
\gamma+m\right]$. The FW field transformation \cite{r5} consists of writing the
equation of motion in the form
\be
i\partial_0\chi(t,p)=\omega\gamma^0\chi(t,p),
\label{e5}
\ee 
where $\omega$ is the energy obtained from the dispersion relation (\ref{e3}),
and $\chi=U\psi$, where $U$ is to be determined. This form of the equation of
motion explicitly shows the evolution of the positive and negative energy
modes, and leads to the Hamiltonian density
$$H_{\rm FW}=\chi^\dag\omega\gamma^0\chi=\omega\bar\chi\chi,$$
which is Hermitian as long as $\omega$ is real. We will use the FW
transformation to map the non-Hermitian Hamiltonian density $H$, with
eigenvalues $\omega$, to the Hermitian Hamiltonian $H_{\rm FW}$, with the same
eigenvalues.

Since the Hamiltonian density $\cH$ has a similar structure as in the Dirac
case, by analogy with the latter, we seek $U$ in the form
\be
U\equiv\exp\left[\theta~\frac{(\vec p-i\vec b)\cdot\vec\gamma}{\sqrt{p^2-b^2}}
\right]=\cos\theta+\frac{\left(\vec p-i\vec b\right)\cdot\vec\gamma}{\sqrt{p^2
-b^2}}\sin\theta,
\label{e6}
\ee
where $\sqrt{p^2-b^2}$ is imaginary in the infrared regime $p^2<b^2$. Unlike the
Dirac case, $U$ is not unitary: $U^\dag(\theta)\ne U(-\theta)=U^{-1}(\theta)$.
In the {\it ansatz} (\ref{e6}), the FW angle $\theta$ must be determined in
order to obtain the equation of motion (\ref{e5}). It is then easy to see that
\bea
U\cH U^{-1}&=&\gamma^0(\vec p+i\vec b)\cdot\vec\gamma\left[\cos(2\theta)-
\frac{m}{\sqrt{p^2-b^2}}\sin(2\theta)\right]\nonumber\\
&&+\gamma^0\left[m\cos(2\theta)+\sqrt{p^2-b^2}\sin(2\theta)\right].\nonumber
\eea

Canceling the term proportional to $\vec\gamma$ requires that
$$\tan(2\theta)=\sqrt{p^2-b^2}/m.$$
For $0\le{\rm Re}\,(2\theta)\le\pi/2$ this implies that
$$\cos(2\theta)=m/\omega\quad{\rm and}\quad\sin(2\theta)=\sqrt{p^2-b^2}/
\omega.$$
This result reduces to the usual definition of the FW angle $\tan(2\theta)=p/m$
in the Dirac case, and in the present case it is then straightforward to check
that one obtains the expected form 
$$U\cH U^{-1}=\gamma^0\left[m\cos(2\theta)+\sqrt{p^2-b^2}\sin(2\theta)\right
]=\omega\gamma^0.$$
It is interesting that even when $b^2\le m^2$ the infrared regime $p^2<b^2$ is
characterized by a purely imaginary FW angle although the energies $\omega$ are
real. The transition between the real and purely imaginary FW angle corresponds
to the limit $p^2\to b^2$, where $\theta\to0$.

\section{Parity-violating mass term}
\label{s3}
Let us examine the FW transformation for a model that was looked at in the
context of $PT$-symmetric theories \cite{r1}, and whose features are similar to
those of the model examined in Sec.~\ref{s2}, although there was no
Lorentz-symmetry violation. The Lagrangian considered in Ref.~\cite{r1} is
\be
{\cal L}=\bar\psi\left(i\slashed\partial-m-\mu\gamma^5\right)\psi,
\label{e7}
\ee
which contains the anti-Hermitian mass term $\mu\bar\psi\gamma^5\psi=-\mu(\bar
\psi\gamma^5\psi)^\dag$. One can check that the corresponding Hamiltonian
density is
$$H=\bar\psi\left(i\vec\gamma\cdot\vec\nabla+m+\mu\gamma^5\right)\psi,$$
and is $\cPT$-even, but $\cP$-odd and $\cT$-odd. While the Hamiltonian is not
Hermitian, the dispersion relation
\be
\omega^2=m^2+p^2-\mu^2
\label{e8}
\ee
shows that the energies are real for any momentum as long as $m^2\ge\mu^2$. The
dispersion relation (\ref{e8}) is the same as that for (\ref{e3}), when we
substitute $b^2\to\mu^2$. As a consequence, we expect the FW transformation to
be similar to that described in the previous section.

The Schr\"odinger form of the equation of motion obtained from the Lagrangian
(\ref{e7}) is
\be
i\partial_0\psi=\cH\psi,
\label{e9}
\ee
where the Hamiltonian operator is $\cH=\gamma^0\left[\vec p\cdot\vec\gamma+m+\mu
\gamma^5\right]$. By analogy with the model (\ref{e2}), we look for a mapping of
the form
\be
U\equiv\exp\left(\theta~\frac{\vec p\cdot\vec\gamma+\mu\gamma^5}{\sqrt{p^2-\mu^2
}}\right)=\cos\theta+\frac{\vec p\cdot\vec\gamma+\mu\gamma^5}{\sqrt{p^2-\mu^2}}
\sin\theta.
\label{e10}
\ee
The parameter $\theta$ should be determined in order to write the equation of
motion in the form
\be
i\partial_0\chi(t,p)=\omega\gamma^0\chi(t,p),
\label{e11}
\ee
where $\chi\equiv U\psi$ and $\omega$ satisfies the dispersion relation
(\ref{e8}). Analysis similar to that described in the previous section
shows that one must choose
\be
\tan(2\theta)=\sqrt{p^2-\mu^2}/m.
\label{e12}
\ee
Hence, the operator (\ref{e10}) maps the non-Hermitian Hamiltonian density $H$
to the FW Hamiltonian density $H_{FW}=\omega\bar\chi\chi$, which is Hermitian as
long as $m^2>\mu^2$, and has the same eigenvalues $\omega$ as $H$. (The present
FW mapping is different from that described in \cite{r1}, where $U$ contained
the non-Hermitian term $\bar\psi\gamma^5\psi$ only, and therefore led to a
different Hermitian Hamiltonian density.)

\subsection{Alternative description}
Even in the situation $m^2>\mu^2$ for which the energies are real, one can
distinguish the UV regime $p^2>\mu^2$, where the angle $\theta$ defined by
(\ref{e12}) is real, from the IR regime $p^2<\mu^2$, where $\theta$ is purely
imaginary. The identity (\ref{e10}) is based on the property that $(\vec p\cdot
\vec\gamma+\mu\gamma^5)^2=\mu^2-p^2$. One may consider instead
\be
U'=\exp\left(\theta'~\frac{\vec p\cdot\vec\gamma+\mu\gamma^5}{\sqrt{\mu^2-p^2}}
\right)=\cosh\theta'+\frac{\vec p\cdot\vec\gamma+\mu\gamma^5}{\sqrt{\mu^2-p^2}}
\sinh\theta'
\label{e13}
\ee
for which the parameter $\theta'$ must be chosen so that
$$\tanh(2\theta')=\sqrt{\mu^2-p^2}/m$$
in order to obtain the FW form (\ref{e11}) of the equation of motion. One can
easily see that the mappings (\ref{e10}) and (\ref{e13}) are equivalent if one
identifies $\theta=i\theta'$, which corresponds to changing the IR regime $p^2<
\mu^2$ to the UV regime $\mu^2<p^2$ in the FW mapping.

Finally, we note that
$$\lim_{p^2\to\mu^2}U=\lim_{p^2\to\mu^2}U'=1+\frac{\vec p\cdot\vec\gamma+\mu
\gamma^5}{2m},$$
so that the apparent singularity when $p^2\to\mu^2$ is actually a smooth limit,
as one would expect from physical grounds, and the FW transformation is not
singular at $p^2=\mu^2$.

\subsection{Conserved current}
The eigenstate solutions of the equation of motion (\ref{e9}) satisfy
$$\psi(t,\vec p)=\exp(-i\cH t)\psi(0,\vec p).$$
Thus,
$$\psi^\dag(t,\vec p)=\psi^\dag(0,\vec p)\exp(i\cH^\dag t),$$
and the corresponding naive probability density does not respect unitarity
because
$$\psi^\dag(t,\vec p)\psi(t,\vec p)=\psi^\dag(0,\vec p)\exp(i{\cal
H}^\dag t)\exp(-i\cH t)\psi(0,\vec p)\ne\psi^\dag(0,\vec p)\psi(0,\vec
p).$$
For consistency, we therefore need to redefine the probability density,
which we write in the form
$$\rho(t)\equiv\psi^\dag(t)A\psi(t),$$
where $A$ is to be determined so that $\rho(t)=\rho(0)$ for eigenstates. This
leads to
$$A\exp(-i\cH t)=\exp(-i\cH^\dag t)A.$$
The latter identity implies that $A\cH=\cH^\dag A$, which is solved by expanding
$A$ in the basis $(1,\vec\gamma,\gamma^5)$. We find that $A=1+\frac{\mu}{m}
\gamma^5$, and the probability density, which respects unitarity, is 
$$\rho\equiv\psi^\dag\left(1+\frac{\mu}{m}\gamma^5\right)\psi.$$

From this result we find the conserved current as follows. For a general
solution to the equation of motion (\ref{e9}) we have
$$\dot\rho=\dot\psi^\dag\left(1+\frac{\mu}{m}\gamma^5\right)\psi+\psi^\dag
\left(1+\frac{\mu}{m}\gamma^5\right)\dot\psi,$$
and according to the same equation of motion 
$$\dot\psi^\dag=\vec\nabla~\bar\psi\cdot\vec\gamma+i\bar\psi(m-\mu\gamma^5)\quad
{\rm and}\quad\dot\psi=\gamma^0\vec\gamma\cdot\vec\nabla\psi-i\gamma^0(m+\mu
\gamma^5)\psi.$$
As a consequence, we see that $\dot\rho=\vec\nabla\cdot\left[\bar\psi~\vec\gamma
\left(1+\frac{\mu}{m}\gamma^5\right)\psi\right]$. Thus, the conserved current is
$$j^\nu=\bar\psi\gamma^\nu\left(1+\frac{\mu}{m}\gamma^5\right)\psi.$$

Finally, we decompose $\psi$ into right- and left-handed components, $\psi=
\psi_{\rm R}+\psi_{\rm L}$, by using the projection operators $\half\left(1\pm
\gamma^5\right)$:
$$\psi_{\rm R}\equiv\half\left(1+\gamma^5\right)\psi,\quad
\psi_{\rm L}\equiv\half\left(1-\gamma^5\right)\psi.$$
Note that $\psi_{\rm R}$ and $\psi_{\rm L}$ are eigenstates of $\gamma^5$
($\gamma^5\psi_{\rm R}=\psi_{\rm R}$, $\quad\gamma^5\psi_{\rm L}=-\psi_{\rm L}$)
and that they are orthogonal (${\psi_{\rm R}}^\dag\psi_{\rm L}={\psi_{\rm
L}}^\dag\psi_{\rm R}=0$). We then observe that the probability density becomes
$$\rho=\left(1+\frac{\mu}{m}\right)|\psi_{\rm R}|^2+\left(1-\frac{\mu}{m}\right)
|\psi_{\rm L}|^2,$$
which is always positive in the relevant regime where $\mu^2\leq m^2$. 
Thus, $\mu=m$ and $\mu=-m$ are interesting special cases in which the
contribution to the density is entirely from right- or left-handed degrees
of freedom.

To conclude, we have discussed two potentially relevant applications of the
non-Hermitian mass term $\mu\gamma^5$. First, because right and left-handed
helicities do not contribute in the same way to the probability density, this
model could shed new light on neutrino physics. A dynamical mechanism to
generate this non-Hermitian mass term is one possible extension of this work.
It would then be interesting to gauge this model and study the effect of the
non-Hermitian mass term on the chiral anomaly, which might be cancelled
independently of the number of lepton generations.


\begin{thebibliography}{99}

\bibitem{r1} C.~M.~Bender, H.~F.~Jones, and R.~J.~Rivers, Phys.~Lett.~B {\bf
625}, 333 (2005). 
 
\bibitem{r2} K.~Jones-Smith and H.~Mathur, Phys.~Rev.~A {\bf 82}, 042101
(2010).

\bibitem{r3} C.~M.~Bender and S.~P.~Klevansky, Phys.~Rev.~ A {\bf 84}, 024102
(2011). 

\bibitem{r4} K.~Jones-Smith and H.~Mathur, Phys.~Rev.~D {\bf 89}, 125014 (2014).

\bibitem{r5} L.~L.~Foldy and S.~A.~Wouthuysen, Phys.~Rev.~{\bf 78}, 29 (1950).

\bibitem{r6} B.~Gon\c{c}alves, M.~M.~D.~J\`unior and B.~J.~Ribeiro, arXiv:1406.5477
 
\bibitem{r7} D.~Colladay and V.~A.~Kostelecky, Phys.~Rev.~D {\bf 58}, 116002
(1998). 
For a review see R.~Bluhm, Lect.~Notes Phys.~{\bf 702}, 191 (2006).
 
\bibitem{r8} J.~Alexandre, Adv.~Math.~Phys.~{\bf 2014}, 527964 (2014).

\bibitem{r9} G.~M.~Shore, Nucl.~Phys.~B {\bf 717}, 86 (2005). 

\bibitem{r10} B.~Altschul, Phys.~Rev.~D {\bf 90}, 021701 (2014).

\bibitem{r11} C.~M.~Bender and M.~Gianfreda, arXiv:1409.3828.

\bibitem{R7} V.~A.~Kostelecky and R.~Lehnert, Phys.~Rev.~D {\bf 63}, 065008
(2001). 

\end{thebibliography}
\end{document}